# Parity Metamaterials and Dynamic Acoustic Mimicry

Jinjie Shi[a,†], Hongchen Chu[b,†], Aurélien Merkel[d], Chenkai Liu[a], Johan Christensen[c,*], Xiaozhou Liu[a,*], Yun Lai[a,*]

[a]MOE Key Laboratory of Modern Acoustics, National Laboratory of Solid State Microstructures, School of Physics, Collaborative Innovation Center of Advanced Microstructures, and Jiangsu Physical Science Research Center, Nanjing University, Nanjing 210093, China
[b]School of Physics and Technology, Nanjing Normal University, Nanjing 210023, China
[c]IMDEA Materials Institute, Calle Eric Kandel, 2, 28906, Getafe, Madrid, Spain
[d]Université de Lorraine, CNRS, Institut Jean Lamour, F-54000 Nancy, France

†These authors contributed equally to this work
*Corresponding authors: Johan Christensen (johan.christensen@imdea.org); Xiaozhou Liu (xzliu@nju.edu.cn); Yun Lai (laiyun@nju.edu.cn)

Abstract: While parity transformation represents a fundamental symmetry operation in physics, its implications remain underexplored in metamaterial science. Here, we introduce a framework leveraging parity transformation to construct parity-inverted counterparts of arbitrary three-dimensional meta-atoms, enabling the creation of parity-engineered metamaterial slabs. We demonstrate that the synergy between reciprocity and parity transformation, distinct from mirror operation, guarantees undistorted wave transmission across exceptional bandwidths, independent of structural configuration or meta-atom design specifics. Furthermore, these metamaterials exhibit dynamic acoustic mimicry capability, enabling adaptive blending of reflected signatures into surrounding environments while preserving transmitted wavefront integrity. Validated through numerical simulations and experimental prototypes, this breakthrough offers transformative potential for acoustic camouflage applications, particularly for sonar systems. Our findings reveal fundamental implications of parity transformation in artificial materials, establishing parity engineering as a paradigm for designing ultra-broadband functional materials with unprecedented operational versatility.

## 1. Introduction

Parity transformation, the symmetry operation [1] involving a change of the sign in all spatial coordinates, i.e., $(x, y, z) \rightarrow (-x, -y, -z)$, has many profound impacts in physics. Applying parity transformation to an arbitrary object creates its unique counterpart as a rotation of its mirror image with reversed chirality, as exemplified by a hand and its image in the mirror in Fig. 1a and b. Recently, parity-time symmetry [2-5] has been exploited to reveal the existence of exceptional points [6-11], opening a promising field of non-Hermitian physics [12]. However, to date, there has been little discussion on the possibility of applying parity transformation alone in acoustic metamaterials [13-20] and metasurfaces [21, 22], which have dramatically broadened the boundaries of acoustic materials over the past decades [23-39].

In this work, we apply parity transformation to design a class of acoustic metamaterials denoted parity metamaterials. Parity transformation relates an arbitrary three-dimensional (3D) meta-atom with its unique parity-inverted counterpart. They constitute a pair of building blocks (Fig. 1c) for a special metamaterial slab called parity metamaterial (Fig. 1e), which can keep transmission wavefront undistorted in an extremely broad spectrum, while dynamically tuning the wavefront in reflection to mimic those from a periodic terrain, a rugged terrain, and a flat terrain, etc. Such metamaterials thus enable dynamic acoustic mimicry (note S1, Supplementary materials) to blend in acoustic environment. At the same time, ultra-broadband undistorted transmission is guaranteed irrespective of the dynamical mimicry, which is extremely important for sonar detection. Therefore, this metamaterial offers a way to realize dynamic acoustic camouflage for advanced sonar systems, just like octopuses that can adapt their color and form to blend in their surroundings but can still clearly see environment (Fig. 1d). Our theory unveils the principle of parity engineering in metamaterials and metasurfaces, promising a profound impact in multiple disciplines.

## 2. Methods

*2.1 Numerical simulations*

The full-wave simulations are performed using the commercial finite element software COMSOL Multiphysics. In the calculations, all solid structures are set as rigid, and the parameters of air are set as $\rho_0 = 1.21$ kg/m$^3$ and $c_0 = 343$ m/s, respectively. The impedance of air is $Z_0 = 415.03$ Pa $*$ s/m$^3$. The resin structures are treated as acoustically rigid materials. The background pressure field is used in Fig. 2 to 4. The periodic boundary condition is set in the *x* and *y* direction and perfectly matched layers are adopted in the *z* direction to reduce the reflection. In Fig. 5, a Gaussian wave is utilized.

*2.2 Experimental measurements*

All samples are fabricated with resin by using stereolithography 3D printing techniques (0.2 mm in precision). All the rotors are printed separately to facilitate rotational operations during the experiment. The whole size of the sample is $360 \times 360 \times 360$ mm$^3$.

The experimental configuration is depicted in Fig. S2 (Supplementary materials). To generate a quasi-plane wave, a speaker array (Five HiVi B1S) equipped with a parabolic mirror is meticulously constructed. A microphone (GRAS 46BE) is affixed to a movable stage to systematically scan the distribution of the acoustic field with a step size of 10 mm. In order to mitigate the influence of waves propagating around the specimen, sound-absorbing foams are affixed in its vicinity. The two measured regions on the sides of reflection and transmission in the *xz*-plane are shown as two blue rectangular areas in Fig. S2a. Both regions are $36 \times 26$ mm$^2$ in size and are positioned 2 cm away from the sample. The photo of the experimental setup is shown in Fig. S2b. The measured reflected field distribution is obtained by subtracting the incident field (measured without the sample) from the total field (measured with the sample). The experiment is carried out in an anechoic room to minimize reflection and noise.

## 3. Results

### 3.1. Theory and design of parity metamaterials

We consider an arbitrary 3D meta-atom and its parity-inverted counterpart, as denoted by P$_1$ and P$_2$, respectively. In order to accord with the working wavelength, the basic units of the metamaterial are composed of an array of 2 × 2 meta-atoms P$_1$ or P$_2$, as depicted in Fig. 1c. The adjacent basic units are separated by hard boundaries to minimize the coupling between them. The length and thickness of the array are specified as $A = 60$ mm and $H = 24$ mm, respectively. To achieve dynamic tunability of the parity metamaterials, we have designed the meta-atom and its parity-inverted counterpart to be rotatable. The magnified view of the inner rotor is displayed in the bottom-right inset of Fig. 1e. The curved plate can be rotated to tune the reflection phase, and the connection to the shaft is set to be asymmetric to remove all symmetry in the meta-structure. The detailed parameters of the meta-atom can be found in note S2 (Supplementary materials). Figure 1e depicts the schematic diagram of the parity metamaterial slab composed of a selected arrangement of the meta-atom and its parity-inverted counterpart, which can tune the reflection while keeping transmission unaffected, if the system is reciprocal.

The underlying physical principle is described as follows. We first consider the scattering properties of this pair of meta-structures (P$_1$ and P$_2$). Assuming that the transmission and reflection coefficients of P$_1$ in Fig. 1f are denoted as *t* and *r*, respectively. When the system is reciprocal, the reciprocity theorem [40-42] asserts that the exchange of the incidence and transmission channels does not change the transmission coefficient, i.e., $t' = t$, as shown in Fig. 1g. On the contrary, the reflection coefficient can differ significantly after the interchange, i.e., $r' \neq r$. Subsequently, we apply a parity operation to the system, as shown in Fig. 1h. P$_1$ is transformed into its counterpart, i.e., P$_2$, which has the same transmission coefficient as P$_1$, i.e., $t'' = t' = t$. This equivalence is regardless of the details of the meta-structures, as well as the frequency and the incident angle. In reflection, contrarily, we have $r'' = r' \neq r$, thus there is a phase difference between $r'$ and *r* over a wide spectrum, which can modulate the reflection from this metamaterial.

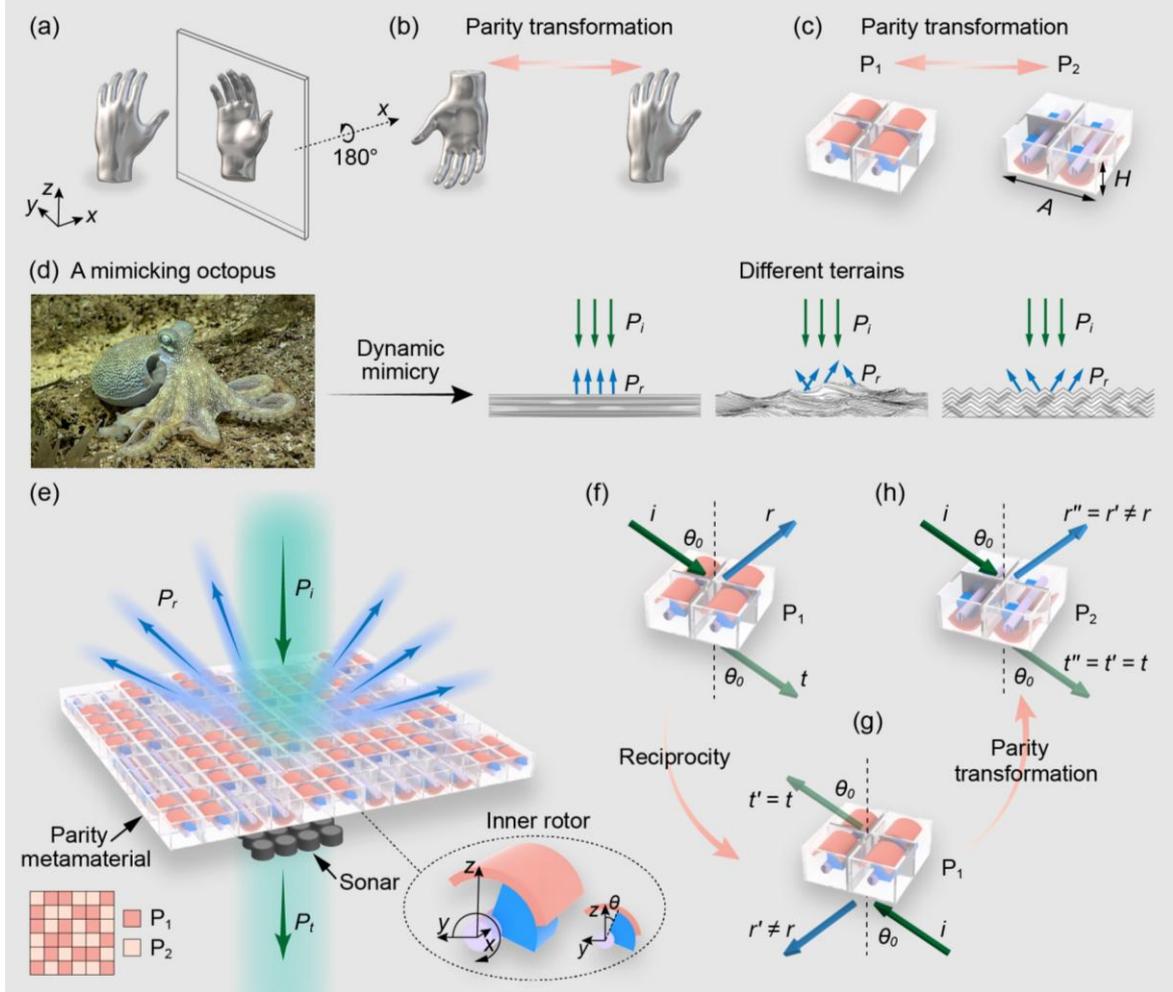

**Fig. 1.** A parity metamaterial composed of a meta-atom and its parity-inverted counterpart. (a) and (b) The hand, its mirror image, and its parity-inverted counterpart. The mirror image and the parity-inverted counterpart are related by a 180° rotation along the *x*-axis. (c) The 3D meta-atom and its parity-inverted counterpart design related by parity transformation. (d) The concept of acoustic mimicry to imitate different terrains, including a flat terrain, a rugged terrain, and a periodic terrain, like an octopus. (e) The illustration of a parity metamaterial based on a random distribution of $P_1$ and $P_2$. Bottom right inset provides a magnified picture of the inner rotor. (f) to (h) Reciprocity and parity transformation rigorously prove that the meta-atom and its parity-inverted counterpart have the same transmission but different reflection coefficients.

*3.2. Ultra-broadband and wide-angle undistorted transmission*

In the following, we present the transmission and reflection properties of $P_1$ and $P_2$. Fig. 2a and b illustrate, irrespectively, the calculated transmittance and transmission phase, and the reflectance and reflection phase of $P_1$ and $P_2$, as functions of the frequency. The direction of the incident wave is along the *z* direction. Here, acoustic dissipation is neglected for simplicity, but this principle also applies to general dissipated systems. From Fig. 2a, it is clearly seen that the transmittance and transmission phase of $P_1$ and $P_2$ are identical over an ultra-broad spectrum ranging from 0.1 to 7 kHz (see note S3, Supplementary materials). The reflectance is also the same, but there is a significant difference in the reflection phase, as depicted in Fig. 2b. Especially, the phase difference, $\Delta\varphi_r$, reaches the maximum value of 180° at a frequency of 5.68 kHz (denoted by a gray vertical line). By rotating $P_1$ and $P_2$ simultaneously, the phase difference in reflection can be conveniently tuned. Figure 2c illustrates the reflection phase difference between $P_1$ and $P_2$, i.e., $\Delta\varphi_r$, as a function of the rotation angle of the rotor and frequency, which covers the whole range of 360° around 5.6 kHz. The black dotted line represents the condition of $\Delta\varphi_r = 180°$. We emphasize that rotation does not change the condition of equal transmittance and $\Delta\varphi_t = 0°$ over the whole spectrum from 0.1 to 7 kHz (see note S4, Supplementary materials). This condition guarantees that the transmitted acoustic wavefront is the same as that of the incidence in an ultra-broad spectrum, independent of the arrangement of $P_1$ and $P_2$.

To demonstrate this unique feature, we consider a parity metamaterial composed of randomly distributed $P_1$ and $P_2$ illuminated by an incident plane wave of different frequencies and incident angles. The arrangement of $P_1$ and $P_2$ is shown in the bottom-left inset of Fig. 1e. By using finite-element software Comsol Multiphysics, the 3D far-field radiation power patterns in transmission are calculated under normal incidence at 2 kHz, 4 kHz, and 6 kHz, as shown in Fig. 2d. Clearly, the direction of transmission ($P_t$) is the same as that of the incidence ($P_i$) for all frequencies. In Fig. 2e, we plot the 3D far-field radiation power patterns in transmission for incident angles of 10°, 30°, and 50° at 4 kHz. It is clearly observed that the direction of transmission ($P_t$) aligns consistently with the direction of incidence ($P_i$) for all angles. These effects are further verified by the calculated near-field distributions, as shown in Fig. 2f and g, respectively. In other words, such a disordered metamaterial functions like an ordered metamaterial in transmission, ensuring its feasibility in important applications like acoustic detection and sonar technology [43-45].

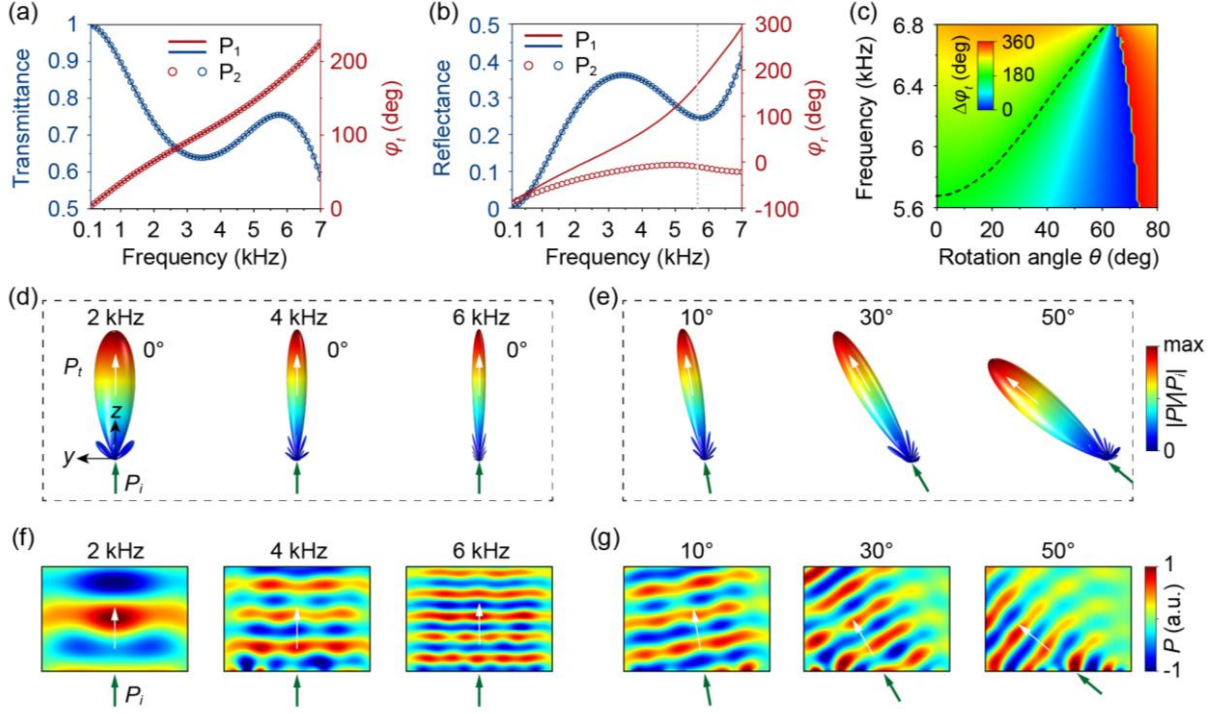

**Fig. 2.** Transmission and reflection properties of $P_1$ and $P_2$. (a) Transmittance and transmission phase spectra of $P_1$ and $P_2$. (b) Reflectance and reflection phase spectra of $P_1$ and $P_2$. (c) Reflection phase difference between $P_1$ and $P_2$ as a function of the rotation angle and frequency. (d) Simulated 3D far-field radiation power patterns in transmission under normal incidence at 2 kHz, 4 kHz, and 6 kHz. (e) Simulated 3D far-field radiation power patterns in transmission under incident angles of 10°, 30°, and 50° at 4 kHz. Each subpanel in (d) and (e) is normalized. (f) and (g) Simulated near-field distributions of the transmitted acoustic field in the $yz$-plane. Here, the colors indicate the normalized pressure of the acoustic field.

*3.3. Parity transformation v.s. mirror operation*

To elucidate the essential role of parity transformation, we compare it with the mirror operation, i.e., upside-down flip, namely $(x, y, z) \rightarrow (x, y, -z)$. When the meta-structure exhibits a $C_4^v$ symmetry in the $xy$-plane, there is no difference between $(x, y, z) \rightarrow (-x, -y, -z)$ and $(x, y, z) \rightarrow (x, y, -z)$ [46, 47]. However, the difference becomes huge when the meta-structure has no symmetry, which is the case here. Figure 3a depicts the schematic diagrams of $P_1$, $P_2$, and the structure generated by mirror operation, $M_z$, respectively. It is seen clearly that $M_z$ is different from $P_2$. In Fig. 3b and c, we plot the transmittance and transmission phase spectra of $P_1$, $P_2$, and $M_z$ under the illumination of an incident angle of 30°, respectively, as shown by the green arrow in Fig. 3a. Clearly, $P_1$ and $P_2$ have the same transmittance and transmission phase, as strictly protected by reciprocity and parity transformation. But $M_z$ and $P_2$ clearly exhibit distinctly different transmission phase around 6.9 kHz. A comprehensive discussion can be found in note S5 (Supplementary materials).

Such a fundamental difference between parity transformation and mirror operation is further experimentally verified. We construct two metamaterials with the sequences $P_1P_2P_1P_2P_1P_2$ (Case I) and $P_1M_zP_1M_zP_1M_z$ (Case II), as illustrated in Fig. 3d and h, respectively. The simulated 3D far-field radiation power patterns in transmission for cases I and case II under an incident angle of 30° at 6.9 kHz are shown in Fig. 3e and i, respectively. Clearly, the transmitted wave ($P_t$) is undistorted in Fig. 3e, but clearly disrupted in Fig. 3i. Such a distinct difference originates in the transmission phase difference between $P_1$ ($P_2$) and $M_z$, which reaches 87° at 6.9 kHz (denoted by a gray vertical line in Fig. 3c). This huge difference is also observed in the calculated and measured near-field distributions for cases I and case II, as shown in Fig. 3f, g, j, and k, respectively. The measured regions are marked by the blue dotted boxes in the $yz$-plane, as shown in Fig. 3f and j. The experimental and numerical results approximately agree with each other, both confirming it is the parity transformation instead of the mirror operation that can preserve the transmission wavefront.

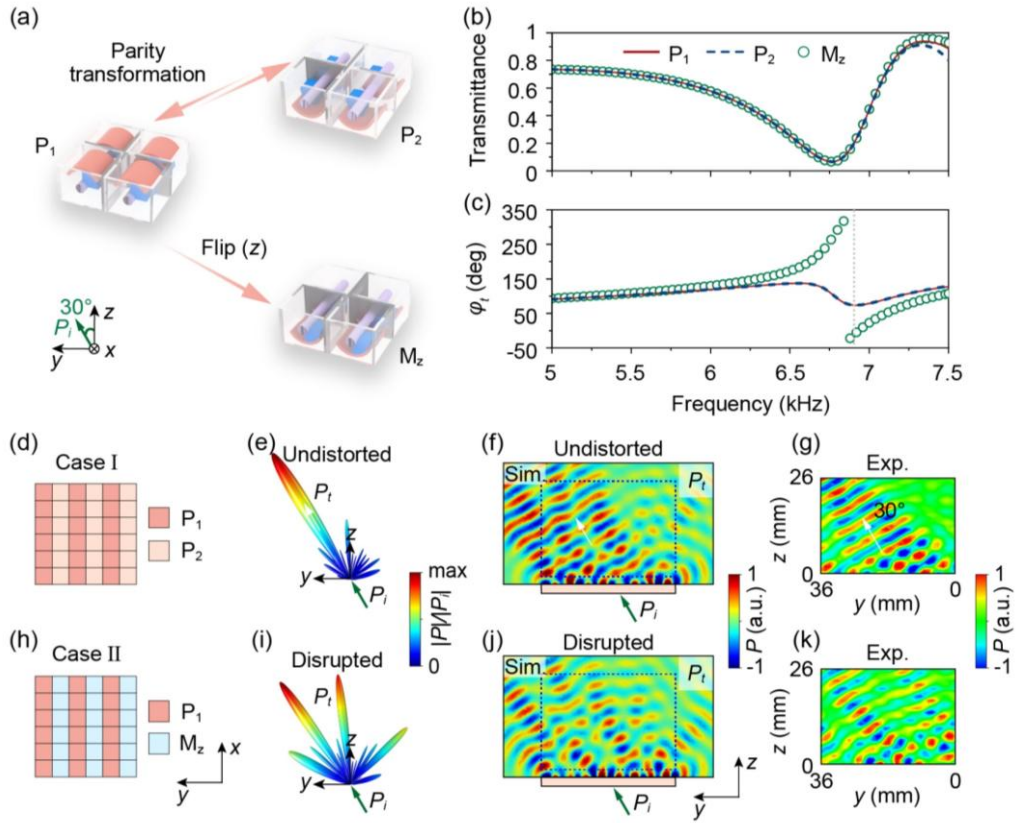

**Fig. 3.** Comparison between parity transformation and mirror operation. (a) Schematic diagrams of $P_1$, $P_2$, and $M_z$. $M_z$ is the mirror image of $P_1$ along the $z$ direction. (b) and (c) show

the transmittance and transmission phase spectra of $P_1$, $P_2$, and $M_z$ under the illumination of an incident angle of 30°. (d) and (h) show the metamaterial design for case I and case II, respectively. (e) and (i) show the simulated 3D far-field radiation power patterns in transmission for case I and case II, respectively. (f), (j), (g), and (k) show the simulated (f, j) and measured (g, k) near-field distributions of the transmitted field for case I (f, g) and case II (j, k) in the *yz*-plane.

*3.4. Dynamic acoustic mimicry*

In the following, we numerically and experimentally demonstrate the realization of dynamic acoustic mimicry. By rotating the rotors of the designed parity metamaterial, it is possible to alter the reflection to emulate that from a periodic terrain, a rugged terrain, and a flat terrain, while keeping the transmission wavefront undistorted. The insets of Fig. 4a portray the magnified views of the inner rotors of $P_1$ and $P_2$, wherein the rotors of both $P_1$ and $P_2$ are simultaneously rotated by the same angle, such that $P_1$ and $P_2$ can always be transformed into each other via parity transformation. Consequently, $P_1$ and $P_2$ always have identical transmittance (see Fig. S7, Supplementary materials) and transmission phase (Fig. 4a), regardless of the rotation angle. On the other hand, the reflection phase difference between $P_1$ and $P_2$ varies significantly with the rotation angle. Figure 4a depicts the calculated reflection phase of $P_1$ and $P_2$, as well as their difference as functions of the rotation angle $\theta$ under normal incidence at a working frequency of 5.68 kHz. The reflection phase difference $\Delta\varphi_r$ varies from -180° to 180°. While the transmission phase difference $\Delta\varphi_t$ remains 0°. This indicates that the rotation angle $\theta$ offers a freedom to dynamically tune the reflection without changing the transmission wavefront.

Here, we demonstrate the switching of reflection from two-beam reflection to specular reflection, while maintaining undistorted transmission all the time. The parity metamaterial is the same as that shown in Fig. 3d. Figure 4b and c show the simulated 3D far-field radiation power patterns under normal incidence at 5.68 kHz for $\theta = 0°$ and $\theta = 73°$ (denoted by gray vertical lines in Fig. 4a), respectively. The direction of transmission ($P_t$) is the same as that of the incidence ($P_i$) for both cases. On the other hand, the reflection changes from the case of splitting into two beams (for $\theta = 0°$) to the case of specular reflection ($\theta = 73°$). This is because $\Delta\varphi_r$ changes from 180° to 0° when $\theta$ changes from 0° to 73°. The switching phenomenon is further verified by simulated and measured near-field distributions, which are shown in Fig. 4d to i, respectively. The scanned areas correspond to the blue dotted boxes in Fig. 4d and e. The measured results coincide excellently with the numerical results, clearly confirming that the transmitted wave ($P_t$) maintains a plane wavefront for the two cases. While for reflection, the angle of reflection is $\theta_r = 30.2°$ ($\theta_r = sin^{-1}(\lambda/(2A))$) (see note S6, Supplementary materials) for rotation angle $\theta = 0°$, as shown by the green arrows in Fig. 4d. Additionally, the case for $\theta = 42°$ is shown in note S6 (Supplementary materials), displaying the simultaneous existence of three-beam reflection and undistorted transmission. Therefore, dynamic acoustic mimicry can be realized via employing the rotation angle $\theta$ as an extra degree of freedom in the meta-structure design.

We have also demonstrated the cases of diffuse reflection and reflection holography (see notes S7 and S8, Supplementary materials). The results again confirm the conclusion: the acoustic reflection signatures can be freely altered to mimic sophisticated acoustic environment, while keeping the transmitted wavefront unchanged in an ultra-broad spectrum, which is crucial for sonar.

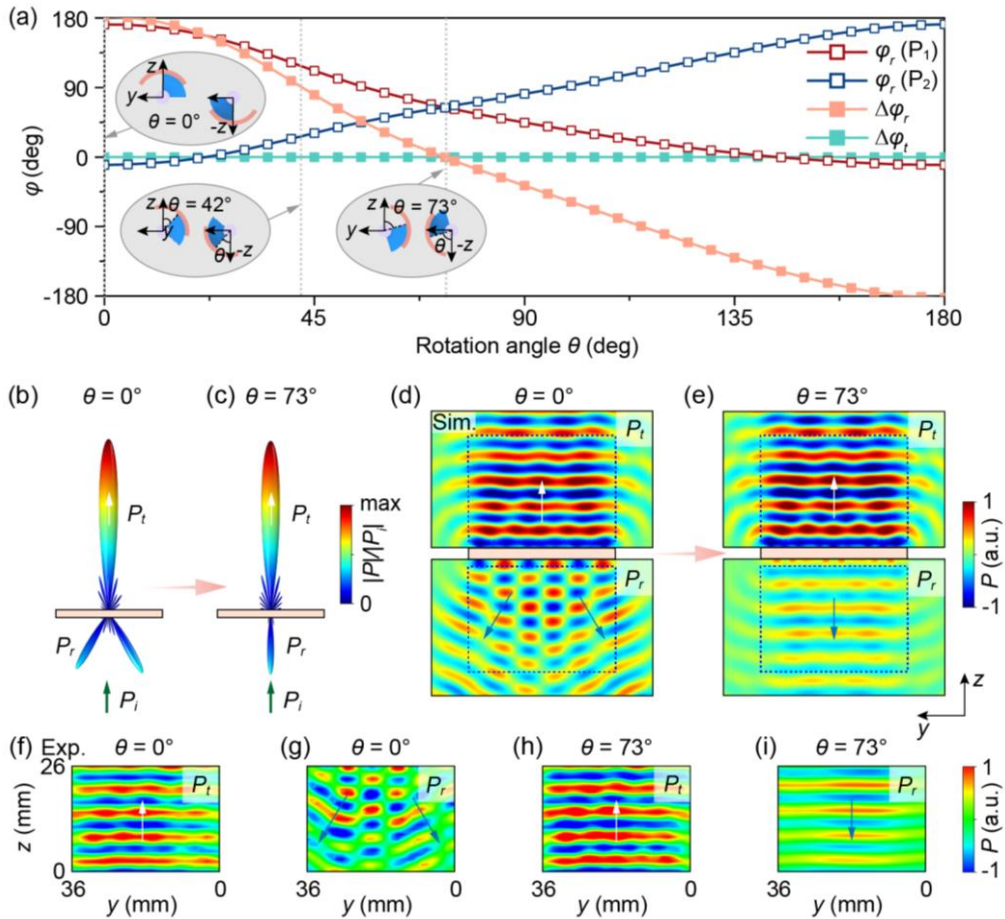

**Fig. 4.** Dynamic acoustic mimicry using a parity metamaterial. (a) Reflection phase spectra of $P_1$ and $P_2$ and their difference in reflection and transmission as functions of the rotation angle $\theta$

at 5.68 kHz. Insets show the magnified views of the inner rotors of P$_1$ and P$_2$. (b) and (c) Simulated 3D far-field radiation power patterns under normal incidence at 5.68 kHz for $\theta = 0°$ and $\theta = 73°$, respectively. (d) and (e) Simulated corresponding near-field distributions in the yz-plane. (f) to (i) Measured acoustic field distributions of transmitted (f, h) and reflected (g, i) waves under normal incidence at 5.68 kHz when $\theta = 0°$ (f, g) and $\theta = 73°$ (h, i).

*3.5 Temporal acoustic camouflage*

The integration of parity metamaterials with sonar systems offers the potential for temporal camouflage. This effect is vividly demonstrated by comparing the reflection signals of a Gaussian pulse incident on a sonar, both without and with a parity metamaterial. The pulse is a time-domain Gaussian signal with a duration of 1 s and a center frequency of 10 kHz. In accordance with the working wavelength, the basic units of the parity metamaterial are composed of only one meta-atom P$_1$ or its parity-inverted counterpart P$_2$, as shown in Fig. 5a. In our simulations, the sonar is modeled as an impedance boundary with an impedance of 38Z$_0$, where Z$_0$ is the impedance of air. Under this condition, the simulated sonar provides approximately 90% reflection.

The parity metamaterial is closely attached to the sonar (just like the necessary sonar domes in practical sonar systems). In this case, the reflections from the metamaterial and the sonar itself both influence the overall reflection. By meticulously adjusting the rotation angle $\theta$ of the rotors in the parity metamaterial, as well as the distance *d* between the metamaterial and the sonar, it is possible to achieve destructive interference between the reflections from the sonar and those from the parity metamaterial. This results in a significant reduction of the overall signal. The mechanism behind the acoustic camouflage is illustrated in Fig. 5b. The total reflection *R* can be expressed as

$$R = r_1 + \sum_{n=1}^{\infty} t_1^2 {r_1'}^{n-1} r_2^n e^{-2in\varphi_d} = r_1 + t_1^2 r_2 e^{-2i\varphi_d}(1 - r_1' r_2 e^{-2i\varphi_d})^{-1}$$

where $r_1$ and $r_1'$ denote the reflection coefficients on the front and back interfaces of the parity metamaterial, $r_2$ is the reflection coefficient of the sonar, $t_1$ is the transmission coefficient of the parity metamaterial, and $\varphi_d$ indicates the phase change through the air layer. By setting $R = 0$, the overall specular signal can be eliminated. At this condition, we only need to consider the 0$^{th}$-order transmission and reflection coefficients of the parity metamaterial and sonar. For brevity, higher-order reflection terms are ignored here as they are significantly smaller. The phase shift can be expressed as $\varphi_d = kd$, where *k* is the wavenumber. For example, when the rotation angle of the rotors is $\theta = 75°$, the distance between the metamaterial and the sonar is calculated as $d = 5.9$ mm. The details are shown in note S9 (Supplementary materials).

Figure 5c shows the snapshots of the incident and reflected Gaussian pulse without the parity metamaterial. As expected, a strong specular reflection is observed when the pulse directly impinges on the sonar. In contrast, when the parity metamaterial covers the sonar, the specular reflection is significantly reduced and the scatterings in other directions emerge, as shown in Fig. 5d. To quantify the temporal camouflage effect, we simulated the reflected signals received at a probe located at a certain distance from the sonar. The results, shown in Fig. 5e, demonstrate a significant reduction in the reflected signal intensity when the parity metamaterial is present. Specifically, the reflected signal strength is reduced to below 7% of its original value, and this value is expected to decrease further at greater distances. This effect makes the sonar system more difficult to be detected. Simultaneously, the parity metamaterial preserves the undistorted ultra-broadband transmission, ensuring the effectiveness of sonar detection.

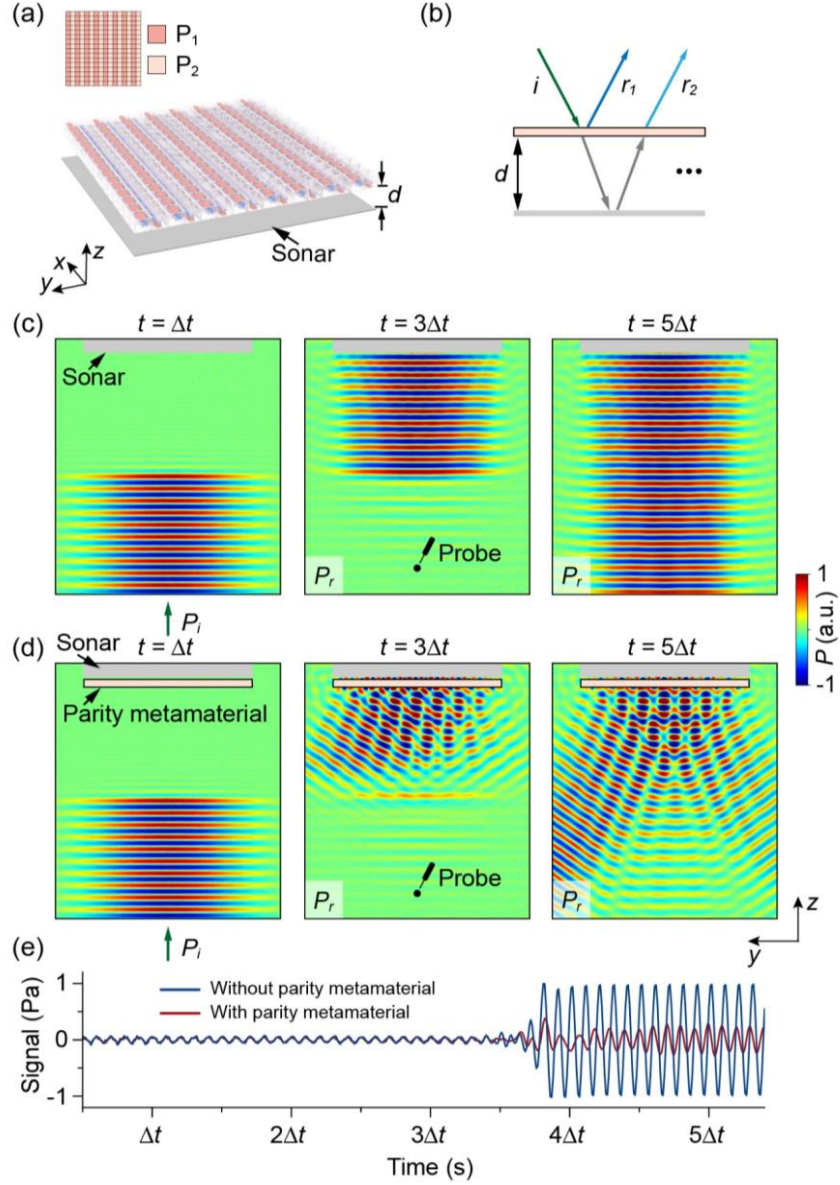

**Fig. 5.** Temporal acoustic camouflage via a sonar integrated with a parity metamaterial. (a) Schematic diagram of a sonar integrated with a parity metamaterial. Inset shows the design of the parity metamaterial. (b) The mechanism of the acoustic camouflage. (c) and (d) Snapshots of a Gaussian pulse incident on a sonar without (c) and with (d) a parity metamaterial, where $\Delta t = 1$ms. The left panels of (c) and (d) show the incident pulse, while the middle and right panels display the reflected signals. (e) Simulated reflection signals received at the probe, both without and with a parity metamaterial.

## 4. Discussion and conclusion

We would like to emphasize that the concept of parity metamaterials is fundamentally different from parity-time (PT) symmetric metamaterials. PT-symmetric metamaterials [2-5] are non-Hermitian systems with global PT-symmetry. While the parity metamaterials do not require any global symmetry at all. On the other hand, PT-symmetric systems require delicately balanced loss and gain. While the functionalities of parity metamaterials are inherently robust to loss (see note S10 in Supplementary materials), because both reciprocity principle and parity transformation are uninfluenced by loss.

Parity metamaterials are also totally different from previous metamaterials studied in topological acoustics [48-51]. The topological properties predominantly rely on the symmetry of the individual unit constituting the topological system. In contrast, parity metamaterials proposed here are constructed using a pair of meta-atoms, i.e. a meta-atom and its unique parity-inverted counterpart. This interplay between the two meta-atoms closely related by parity transformation opens up new possibilities for unprecedented properties.

We should note that parity metamaterials reveal the fundamental difference between parity transformation $(x, y, z) \to (-x, -y, -z)$ and mirror operation $(x, y, z) \to (x, y, -z)$, which was applied in previous optical metasurface designs [46, 47]. This fundamental difference is manifested by the significant advantage that parity metamaterials can be constructed from arbitrary building blocks, including ones with chirality and no symmetry at all, which is a huge advance in contrast to the previous optical designs. This difference also releases a large amount of new freedom that enables the dynamical acoustic mimicry, which can flexibly simulate the acoustic signatures of a periodic terrain, a rugged terrain, and a flat terrain, etc. Furthermore, the acoustic camouflage performance is also demonstrated in temporal domains.

In summary, we introduce the fundamental symmetry operation, parity transformation, to design a new class of metamaterials denoted as parity metamaterials. Such metamaterials are constituted by an arbitrary 3D meta-atom and its unique parity-inverted counterpart. The combination of parity transformation and reciprocity allows for dynamic acoustic mimicry in reflection without distorting the transmitted wavefront across a wide spectrum. This approach is universal and applies to all types of structures and materials, as long as they are reciprocal. The constituent meta-atoms do not necessitate any specific symmetry, and work robustly with loss. Although the demonstration here is conducted using airborne sound, it can also be extended to underwater scenarios (note S11 in Supplementary materials). It is also effective when the thickness of the metamaterial is much larger (note S12, Supplementary materials). Our findings uncover the profound role of parity transformation in ultra-broadband wave manipulation, establishing parity engineering as a transformative paradigm for artificial material design, with implications spanning acoustic camouflage, adaptive metasurfaces, and next-generation communication systems.

**Conflict of Interest**

The authors declare that they have no conflict of interest.

**Acknowledgments**

Yun Lai thanks the support from the National Natural Science Foundation of China (12474293 and 12174188) and the Natural Science Foundation of Jiangsu Province (BK20233001). Xiaozhou Liu acknowledges the support from the National Key R&D Program of China (2020YFA0211400), the National Natural Science Foundation of China (12174192), and the State Key Program of National Natural Science Foundation of China (11834008). Jinjie Shi thanks the support from the China Postdoctoral Science Foundation (2023M731612). Hongchen Chu acknowledges the support from the National Natural Science Foundation of China (12404364) and the Natural Science Foundation of Jiangsu Province (BK20240575).

**Author Contributions**

Yun Lai and Hongchen Chu conceived the idea. Jinjie Shi and Hongchen Chu conducted the analysis, simulations, and sample fabrication. Jinjie Shi conducted the experiment. Aurélien Merkel helped in the theoretical analysis and Chenkai Liu helped in the experiments. Yun Lai, Xiaozhou Liu and Johan Christensen organized and led the project. All the authors contributed to the data analysis and manuscript preparation.

**Appendix A. Supplementary materials**

Supplementary materials to this article can be found online at.

Supplementary materials

# Parity Metamaterials and Dynamic Acoustic Mimicry

Jinjie Shi[a,†], Hongchen Chu[b,†], Aurélien Merkel[d], Chenkai Liu[a], Johan Christensen[c,*], Xiaozhou Liu[a,*], Yun Lai[a,*]

[a]MOE Key Laboratory of Modern Acoustics, National Laboratory of Solid State Microstructures, School of Physics, Collaborative Innovation Center of Advanced Microstructures, and Jiangsu Physical Science Research Center, Nanjing University, Nanjing 210093, China
[b]School of Physics and Technology, Nanjing Normal University, Nanjing 210023, China
[c]IMDEA Materials Institute, Calle Eric Kandel, 2, 28906, Getafe, Madrid, Spain
[d]Université de Lorraine, CNRS, Institut Jean Lamour, F-54000 Nancy, France

†These authors contributed equally to this work
*Corresponding authors: Johan Christensen (johan.christensen@imdea.org); Xiaozhou Liu (xzliu@nju.edu.cn); Yun Lai (laiyun@nju.edu.cn)

**Note S1. Camouflaged sonar dome based on parity metamaterials**

Sonar domes are commonly employed alongside sonar systems and are widely utilized in underwater vessels. Their primary purpose is to house and protect electronic equipment used for detection, navigation, and ranging. To ensure the effectiveness of sonar detection, sonar domes must possess ultra-broadband undistorted acoustic transmission, which is satisfied by the unique function of the parity metamaterials proposed here. Interestingly, the dynamic acoustic mimicry of parity metamaterials is capable of changing the signature of sonar in reflection. When the sonar dome introduces diffuse reflection (depicted by blue color) or other signatures, while preserving the undistorted ultra-broadband transmission (depicted by red color), as shown in Fig. S1, the signature of sonar in reflection is changed, making the sonar system more difficult to be detected by other sonars.

**Note S2. Detailed parameters of the meta-atom**

The diagram of P$_1$ is shown in Fig. S3a, which is the 2 × 2 array of the meta-atom. The length and thickness of the meta-atom are $a = 30$ mm and $H = 24$ mm, respectively. Particularly, the meta-atom is ingeniously designed to be a rotatable rotor. For clarity, Fig. S3b shows the front view, right view, and top view of the meta-atom. The curved plate (the pink part) has a span of $\theta_2 = 120°$ whereby the inner and outer radius of the curved plate are $R_1 = 10$ mm and $R_2 = 12$ mm, respectively. The depth of the curved plate is $d_2 = 22$ mm. The fan structure (the blue part) has a span of $\theta_3 = 90°$ and the thickness is $d_1 = 11$ mm. The middle shaft (the purple part) exhibits a radius of $r = 3$ mm. The thickness of all the hard walls is $t = 2$ mm. Here, the curved plate is primarily used to reflect sound waves, and it can adjust the reflection phase during rotation. The depth can also be changed to tune the transmittance. The shaft provides the meta-atom ("inverse-meta-atom") with rotational freedom. The fan structure connects the curved plate and the shaft, and is set to be asymmetric to remove all symmetry of the meta-atom and "inverse-meta-atom."

**Note S3. Discussion on the strength of sonar signal acquisition**

The strength of the signals collected by the sonar is primarily related to two factors. On one hand, the transmittance of P$_1$ (P$_2$) can be adjusted by the depth of the curved plate, i.e., $d_2$. Figure S4 shows the transmittance of P$_1$ (P$_2$) as a function of the depth of the curved plate at the condition of $\Delta\varphi_r = 180°$. Clearly, the transmittance can be flexibly tuned by changing the depth of the curved plate and can even exceed 95%. On the other hand, the sonar's acquisition intensity can also be enhanced by increasing its sensitivity.

**Note S4. Transmission phase difference between P₁ and P₂**

Figure S5 shows the calculated transmission phase difference between P₁ and P₂ as a function of the rotation angle of the rotor and frequency, demonstrating that the transmission phase difference between P₁ and P₂ is zero over an exceptionally broad spectrum spanning from 0.1 to 7 kHz and across wide angles from 0° to 180°.

**Note S5. Simulated sound pressure distributions of P₁ and M_z**

To demonstrate the difference in transmission between P₁ and M_z, the simulated total sound pressure distributions of P₁ and M_z are shown in Fig. S6. The background pressure fields with incident angles of 30° at 6.9 kHz are set below. The top and bottom are set as perfectly matched layers, where the red lines represent periodic boundary conditions. Owing to inherent structural disparities, the excitation modes inherently diverge, ultimately yielding noteworthy discrepancies in the total sound pressure distributions between P₁ and M_z, as shown in the rectangular dotted boxes of Fig. S6. Consequently, the transmission phase between P₁ and M_z is anticipated to exhibit divergence as well.

**Note S6. Multiple-beam reflection with undistorted transmission**

The diffraction grating equation describes the angles at which acoustic waves of a particular wavelength will be diffracted when they pass through a diffraction grating. The equation is:

$$\sin\theta_m = \frac{m}{D}\lambda \qquad (1)$$

Where $\theta_m$ is the diffraction angle, $\lambda$ is the wavelength of the incident wave, $D$ is the length of the supercell ($D = 2A = 12\ cm$ in our case), and $m$ is the diffraction order (an integer). In Equation (1), to fulfill the requirement $|sin\theta_m| \leq 1$, $m$ should be taken as 0, $\pm 1$ ($\lambda = 6.04\ cm$). When $m = 0$, a phase difference of $\pi$ between the meta-atom and "inverse-meta-atom" leads to vanishing reflection. Consequently, the reflection by the parity metamaterial has only two identical beams ($m = \pm 1$), as shown in Fig. 4d.

It is noted that the reflection phase difference between P₁ and P₂ varies significantly with the rotation angle, covering -180° to 180°. Next, we demonstrate the simultaneous existence of three-beam reflection and undistorted transmission for the rotation angle of $\theta = 42°$. The metamaterial design is shown in Fig. S8a. At this time, the reflection phase difference between P₁ and P₂ becomes 90°. Therefore, the acoustic path difference is $A\ sin\theta_m + \frac{\lambda}{4}$. From Equation (1), the acoustic path difference between P₁ and P₂ within a unit cell is $\frac{(2m+1)\lambda}{4}$. When $m =$ $0, \pm 1$, the phase difference between P₁ and P₂ are $\pm\frac{\lambda}{4}$ and $\frac{3\lambda}{4}$. Therefore, the three beams can coexist on the side of reflection. Fig. S8b shows the simulated 3D far-field radiation power pattern under normal incidence at 5.68 kHz when $\theta = 42°$. The phenomena of undistorted transmission and three-beam reflection can be observed simultaneously.

**Note S7. Diffuse reflection with undistorted transmission**

We demonstrate the case of diffuse reflection and undistorted transmission by designing a parity metamaterial with random configurations, as shown in Fig. S9a. When the rotation angle of P₁ (P₂) is $\theta = 0°$, the reflection phase difference between P₁ and P₂ is 180°. We plot the simulated 3D far-field radiation power pattern under normal incidence at 5.68 kHz, as shown in Fig. S9b. A prominent radiation lobe is clearly seen in the direction of incidence, which corresponds to the transmission ($P_t$) through the parity metamaterial. However, the reflection lobes ($P_r$) of the parity metamaterial spread in many different directions. For clarity, the far-field reflection pattern in a logarithmic coordinate is also shown in the inset of Fig. S9b. It is clearly seen that the reflected-wave energy is distributed uniformly, which is the characteristic feature of diffuse reflection. Diffuse reflection is also one of the most desired functions for a sonar dome, besides the basic requirement of ultra-broadband undistorted transmitted wavefront for detection.

**Note S8. Reflection holography with undistorted transmission**

Here, we also demonstrate the case of reflection holography and undistorted transmission by the parity metamaterial. To enhance imaging accuracy, we design a larger parity metamaterial with 30 × 30 units. The metamaterial design for holography is shown in Fig. S10a, which is obtained by the angular spectrum method.[S1] The target image is shown in Fig. S10b, with the target focal length set to $f_d = 20A = 1.2$ m. In Fig. S10d, we plot the simulated near-field distribution of the transmitted acoustic field under normal incidence at 5.68 kHz. Clearly, the direction of transmission ($P_t$) is the same as that of the incidence ($P_i$). In Fig. S10e, we plot the simulated intensity profile of the reflected wave, where the letter 'U' is clearly visible. We emphasize that the imaging accuracy can be further improved by increasing the size of the parity metamaterial.

**Note S9. The mechanism of the acoustic camouflage**

We consider a scenario where a parity metamaterial is positioned at a distance $d$ from a sonar. When an incident wave interacts with the parity metamaterial, a portion of the wave is reflected (characterized by the reflection coefficient $r_1$), while the remaining part propagates through the parity metamaterial (described by the transmission coefficient $t_1$), as shown in Fig. S11. As the transmitted wave traverses the air layer, it undergoes a phase shift, resulting in a transmission coefficient of $t_1 e^{-i\varphi_d}$, where $\varphi_d$ represents the phase change through the air layer. The wave reflected back from the sonar is characterized by the coefficient $t_1 r_2 e^{-i\varphi_d}$, where $r_2$ is the reflection coefficient of the sonar. Consequently, the wave returning to the parity metamaterial carries a coefficient of $t_1 r_2 e^{-2i\varphi_d}$. This wave is subsequently divided into two components: the reflected wave and the transmitted wave, with coefficients of $t_1 r_1' r_2 e^{-2i\varphi_d}$ and $t_1^2 r_2 e^{-2i\varphi_d}$, respectively. It should be noted that $r_1'$ denote the reflection coefficient on the back interface of the parity metamaterial. These scattering cascades occur infinitely. The total reflection $R$ can be expressed as

$$R = r_1 + \sum_{n=1}^{\infty} t_1^2 r_1'^{n-1} r_2^n e^{-2in\varphi_d} = r_1 + t_1^2 r_2 e^{-2i\varphi_d}(1 - r_1' r_2 e^{-2i\varphi_d})^{-1} \quad (2)$$

Here, we consider the case of normal incidence. For brevity, higher-order terms in reflection are ignored here because they are much smaller. By setting $R = 0$, the overall specular signal can be eliminated. At this condition, we only need to consider the 0$^{th}$-order transmission and reflection coefficients of the parity metamaterial and sonar, which can be computed using the commercial finite element software COMSOL Multiphysics. For example, when the rotation angle of the rotors is $\theta = 75°$, the calculated transmission and reflection coefficients are $t_1 = -0.17569 - 0.20617i$, $r_1 = -0.28323 + 0.434901i$, $r_1' = -0.28334 + 0.43574i$, and $r_2 = 1.1289 - 0.11148i$, respectively. Substituting these values into simplified Equation (2) yields a series of values for $d$. We have selected an appropriate value of $d = 5.9$ mm.

**Note S10. Meta-atom and "inverse-meta-atom" with loss**

We emphasize that the design strategy presented in this work is universal, even in the presence of loss. We take $P_1$ and $P_2$ from Fig. 1 in the main text as an example. Here, the dissipation term here is chosen as 10% loss in the bulk modulus of air. Figure S12a and b show, irrespectively, the calculated transmittance and transmission phase, and the reflectance and reflection phase of $P_1$ and $P_2$, as functions of the frequency. Clearly, even in the presence of loss, $P_1$ and $P_2$ have identical transmittance and transmission phase over an ultra-broad spectrum ranging from 0.1 to 7 kHz. This characteristic ensures that the parity metamaterials composed of $P_1$ and $P_2$ can manifest undistorted transmission wavefront. On the other hand, $P_1$ and $P_2$ exhibit slight differences in reflectance, but there is a notable distinction in the reflection phase. Such a characteristic satisfies the requirement for designing dynamic reflection functionality. Therefore, our design strategy is universally applicable.

**Note S11. Performance of the parity metamaterial underwater**

Here, taking account of the acoustic-elastic coupling, we investigate the performance of the parity metamaterial underwater. We consider the parity metamaterial with a random distribution of P$_1$ and P$_2$ underwater, as shown in Fig. S13a. In underwater applications, the material for parity metamaterial can be chosen to be steel. The parameters of steel are taken to be mass density $\rho = 7850 \, kg/m^3$, the Young's modulus $E = 180 \, Gpa$, and Poisson ratio $\upsilon = 0.25$. The parameters of water are set as $\rho_0 = 1000 \, kg/m^3$ and $c_0 = 1531 \, m/s$. Figure S13a and b illustrate, irrespectively, the calculated transmittance and transmission phase, and the reflectance and reflection phase of P$_1$ and P$_2$, as functions of the frequency. Clearly, even underwater, P$_1$ and P$_2$ have identical transmittance and transmission phase over an ultra-broad spectrum ranging from 26 to 32 kHz. This characteristic ensures that the parity metamaterials composed of P$_1$ and P$_2$ can manifest undistorted transmission wavefront. On the other hand, P$_1$ and P$_2$ exhibit slight differences in reflectance, but there is a notable distinction in the reflection phase. The reflection phase difference between P$_1$ and P$_2$ can reach 180° at 28.95 kHz. Such a characteristic satisfies the requirement for designing dynamic reflection functionality. Figure S13d and e show the simulated 3D far-field radiation power pattern and corresponding near-field distribution in the *yz*-plane under normal incidence. The phenomena of undistorted transmission and diffuse reflection are clearly observed. Therefore, our design strategy is applicable underwater.

**Note S12. Parity metamaterials with richer degrees of freedom**

Interestingly, when the thickness increases, more degrees of freedom could be introduced into the parity metamaterials, thereby enabling more ways of regulation. Here, we demonstrate a meta-atom with four different inclusions, i.e., a sphere, two square panels, and an oblique circular panel (angle *θ$_1$*), distributed vertically, denoted as N$_1$. Its "inverse-meta-atom", N$_2$, is obtained by applying parity transformation to N$_1$, as depicted in Fig. S14a. The adjacent meta-atoms are also separated by hard boundaries. The length and thickness of the meta-atom ("inverse-meta-atoms") are specified as $a = 30$ mm and $H = 140$ mm, respectively. The square panels have cross sections of $18 \times 18$ cm$^2$. The distances are $L_1 = 68$ mm, $L_2 = 15$ mm, and $L_3 = 120$ mm (the distance between the sphere and the circular panel). The centers of the sphere and the circular panel, as well as the center of the meta-atom, lie on the same straight line. The edge of the sphere is aligned with the upper surface of the meta-atom. The radius of the sphere and cylinder is $r_1 = 10$ mm and $r_2 = 12$ mm, respectively. The length and width of the two middle square panels are both $\omega_1 = \omega_2 = 18$ mm. The thickness of the hard walls, square panels, and circular panel are all $t = 2$ mm. The insets of Fig. S14a show the top view of the square panel and the front view of the circular panel, respectively. The angle between the circular panel and the *y*-axis is $\theta_1 = 45°$. We plot the calculated transmission phase difference $\Delta\varphi_t$ and reflection phase difference $\Delta\varphi_r$ as functions of the incident angle and frequency in Fig. S14b and c. It is seen that $\Delta\varphi_t$ is zero over an exceptionally broad spectrum spanning from 0.1 to 6 kHz and across wide angles from 0° to 50°. Conversely, due to the larger thickness, $\Delta\varphi_r$ undergoes multiple transitions from 0° to 360°. At 4.393 kHz, the reflection phase difference precisely reaches 180° (denoted by a black star in Fig. S14c).

Next, we demonstrate that this parity metamaterial of greater thickness exhibits richer degrees of freedom to modulate the reflection. The degrees of freedom we considered are the distances $L_1$, $L_2$, $L_3$, and the angle of the oblique circular panel $\theta_1$. In Fig. S14d, we plot the calculated $\Delta\varphi_t$ and $\Delta\varphi_r$ as functions of $L_1$ and $L_2$ under the normal incidence at 4.393 kHz. It is observed that $\varphi_t = 0$ regardless of the parameters $L_1$ and $L_2$. However, $\Delta\varphi_r$ varies with $L_1$ and $L_2$, covering the whole region from 0° to 360°. Interestingly, the region between the two square panels forms a resonant cavity of length $L_2$, as shown in the inset in Fig. S14a (obtained at parameters marked by a black star in Fig. S14d). We note that $L_1 + L_2$ has a maximum value. In Fig. S14e and f, we plot the calculated $\Delta\varphi_t$ and $\Delta\varphi_r$ as functions of $L_3$ and $\theta_1$. Similarly, $\Delta\varphi_t = 0$ regardless of the parameters $L_1$ and $L_2$. On the contrary, $\Delta\varphi_r$ varies significantly by adjusting $L_3$ and $\theta_1$, covering the whole region from 0° to 360°. This case shows that our design strategy is universal and applicable to general acoustic metamaterials.

[S1] J. W. Goodman, Introduction to Fourier Optics, 2017, 4th ed.

**Figures**

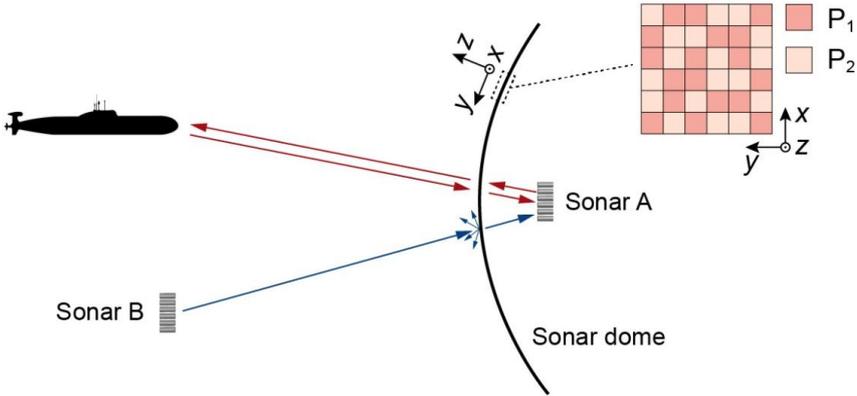

**Fig. S1.** Schematic of the camouflaged sonar dome based on parity metamaterials.

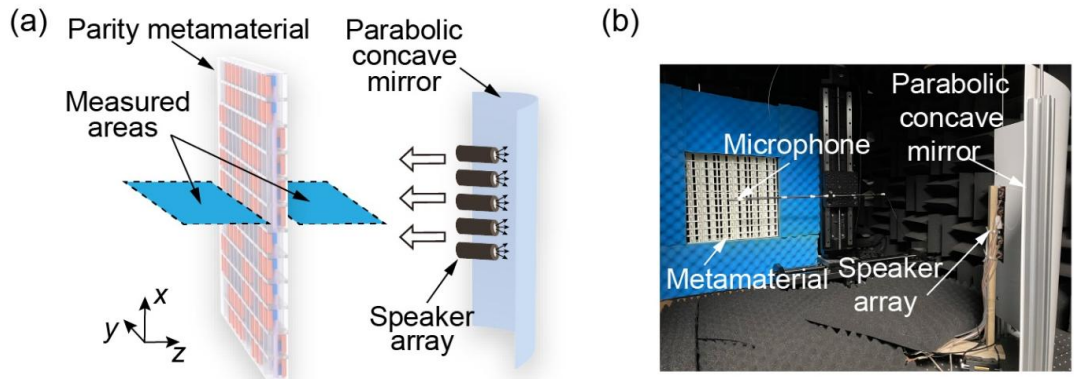

**Fig. S2.** Relevant experimental settings. (a) Schematic diagram of the experimental setup. (b) Photograph of the experimental setup.

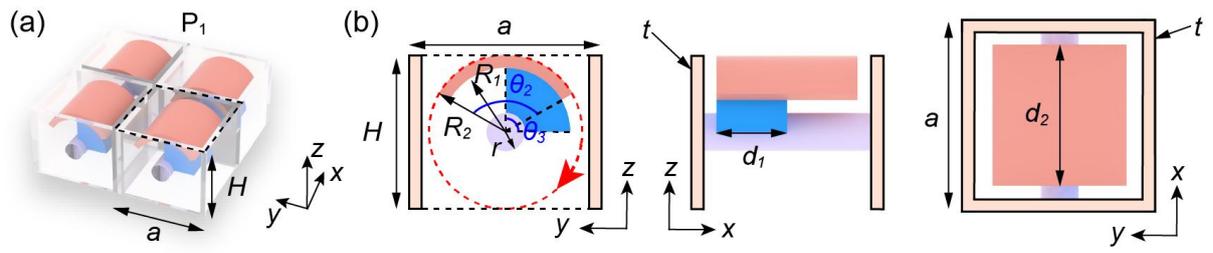

**Fig. S3.** Detailed parameters of the meta-atom. (a) Two-dimensional illustration of $P_1$. (b) Front view, right view, and top view of the meta-atom.

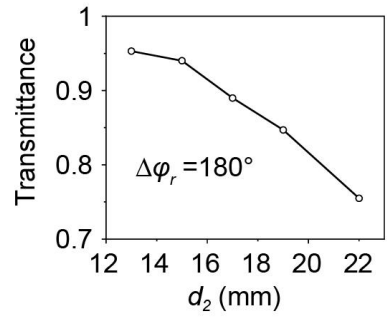

**Fig. S4.** Transmittance of $P_1$ ($P_2$) at the condition of reflection phase difference, $\Delta\varphi_r = 180°$, as a function of the depth of the curved plate, $d_2$.

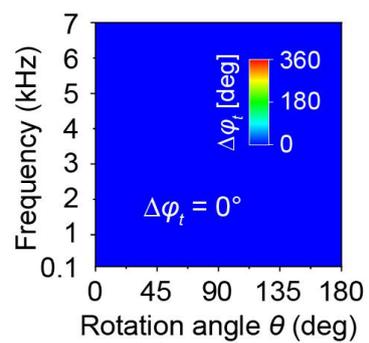

**Fig. S5.** Transmission phase difference between P$_1$ and P$_2$ as a function of the rotation angle of the rotor and frequency.

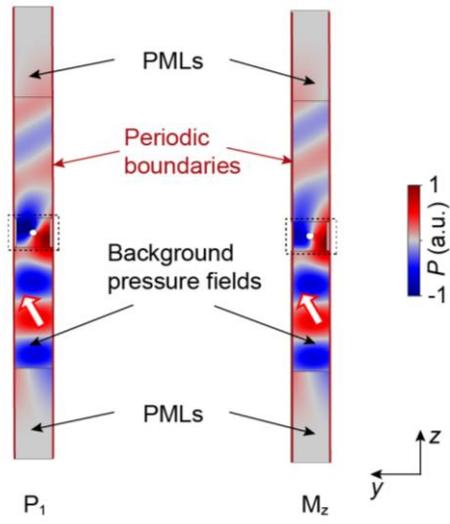

**Fig. S6.** Simulated sound pressure distributions of $P_1$ and $M_z$.

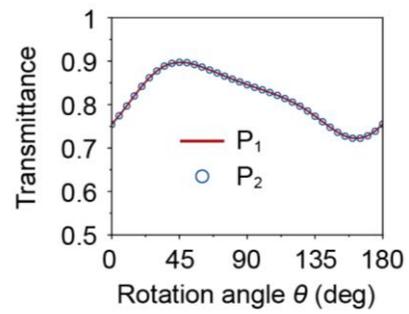

**Fig. S7.** Calculated transmittance of $P_1$ and $P_2$ regarding the rotation angle of the inner rotors at 5.68 kHz. $P_1$ and $P_2$ have identical transmittance due to reciprocity and parity transformation.

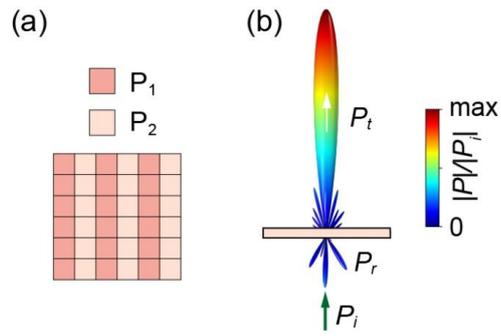

**Fig. S8.** Three-beam reflection with undistorted transmission. (a) The metamaterial design for three-beam reflection. (b) Simulated 3D far-field radiation power pattern under normal incidence at 5.68 kHz when $\theta = 42°$.

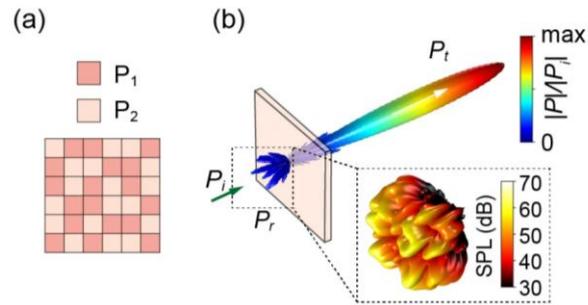

**Fig. S9.** Diffuse reflection with undistorted transmission. (a) The metamaterial design for diffuse reflection. (b) Simulated 3D far-field radiation power pattern under normal incidence at 5.68 kHz. The inset shows the far-field reflection patterns in a logarithmic coordinate.

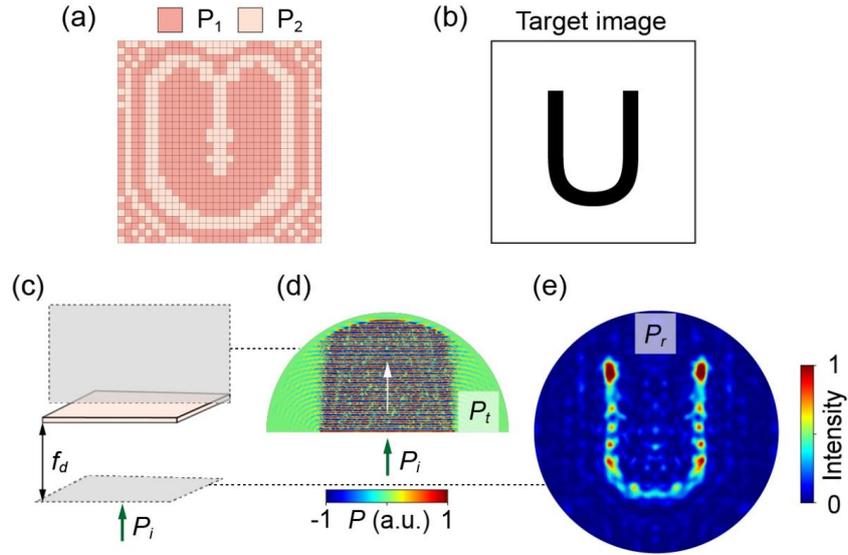

**Fig. S10.** Reflection holography with undistorted transmission. (a) The metamaterial design for holography. (b) Target image. (c) to (e) Simulated near-field distribution of the transmitted acoustic field (d) and the intensity profile of the reflected wave (e) under normal incidence at 5.68 kHz. The grey rectangular boxes correspond to the planes of the simulated near-field results.

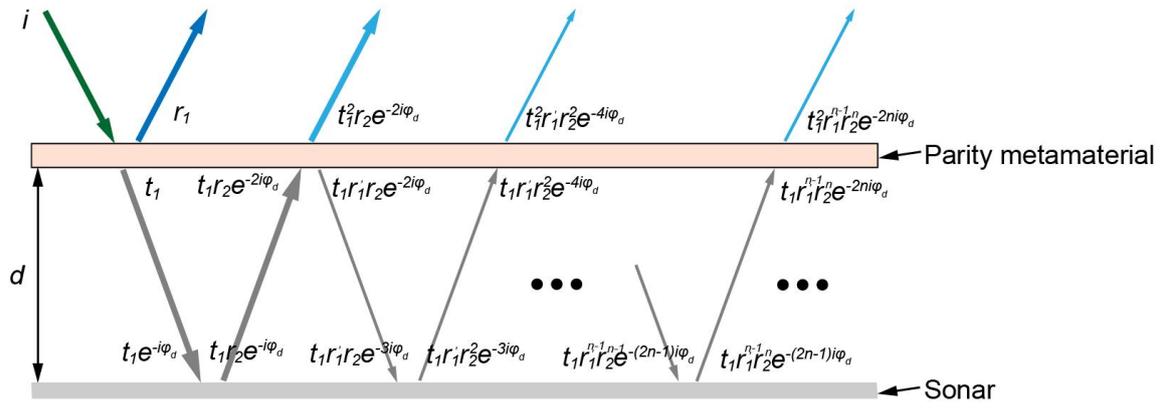

**Fig. S11.** Illustration of the mechanism of the acoustic camouflage.

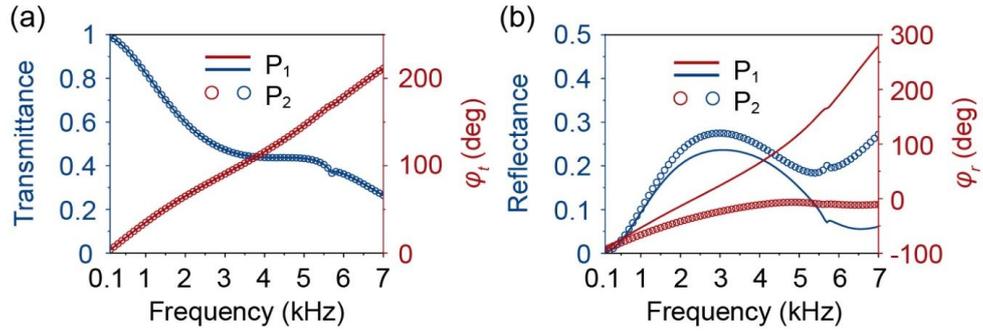

**Fig. S12.** Meta-atom and "inverse-meta-atom" with loss. (a) Transmittance and transmission phase spectra of $P_1$ and $P_2$. (b) Reflectance and reflection phase spectra of $P_1$ and $P_2$.

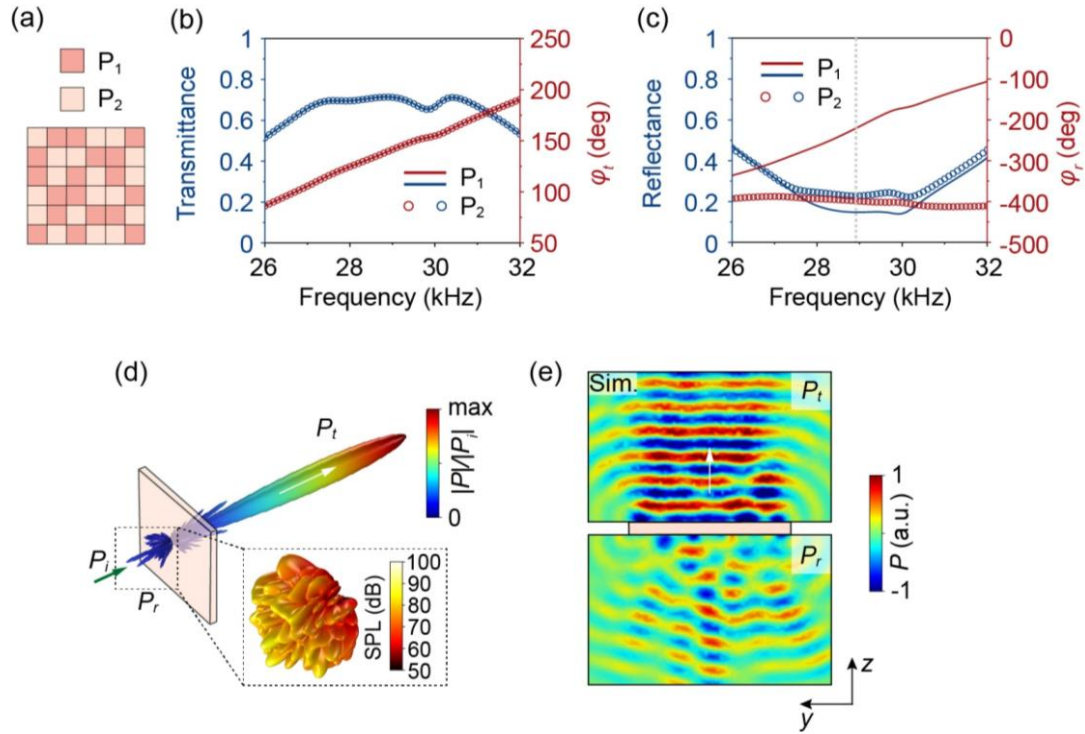

**Fig. S13.** Performance of the parity metamaterial underwater. (a) The design of the parity metamaterial underwater. (b) Transmittance and transmission phase spectra of $P_1$ and $P_2$. (c) Reflectance and reflection phase spectra of $P_1$ and $P_2$. (d) Simulated 3D far-field radiation power pattern under normal incidence at 28.95 kHz. (e) Simulated corresponding near-field distribution in the *yz*-plane.

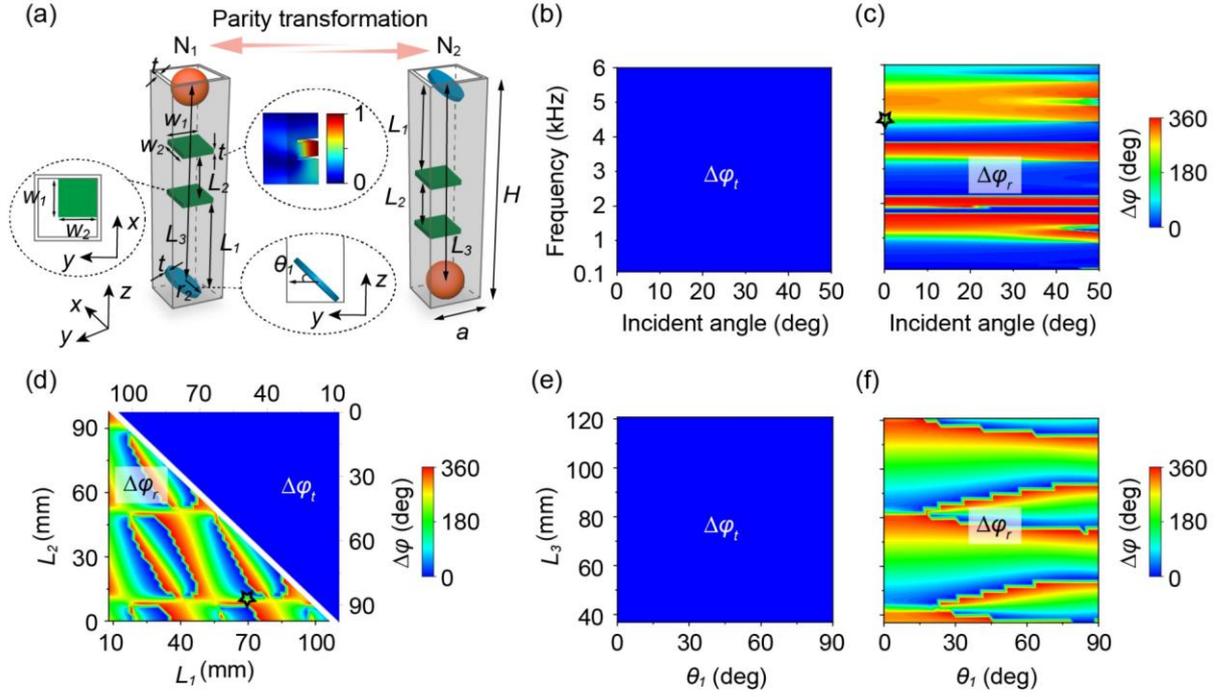

**Fig. S14.** Universality of the design strategy and more degrees of freedom. (a) Schematic diagrams of the meta-atom and "inverse-meta-atom" composed of four different inclusions. The upper inset displays a selected acoustic intensity profile of $N_1$. The lower inset displays the side view of the circular panel. (b) and (c) Calculated transmission phase difference $\Delta\varphi_t$ and reflection phase difference $\Delta\varphi_r$ as functions of the incident angle and frequency. (d) Calculated transmission phase difference $\Delta\varphi_t$ and reflection phase difference $\Delta\varphi_r$ as functions of $L_1$ and $L_2$. (e) and (f) Calculated transmission phase difference $\Delta\varphi_t$ and reflection phase difference $\Delta\varphi_r$ as functions of $\theta_1$ and $L_3$ (with $L_1 = 22$ mm and $L_2 = 10$ mm). The considered frequency is 4.393 kHz.